\documentclass[lettersize,journal]{IEEEtran}
\usepackage{amsmath,amsfonts}
\usepackage{algorithm}
\usepackage{algpseudocode}
\usepackage{amsmath}
\usepackage{array}
\usepackage[caption=false,font=normalsize,labelfont=sf,textfont=sf]{subfig}
\usepackage{textcomp}
\usepackage{stfloats}
\usepackage{url}
\usepackage{verbatim}
\usepackage{graphicx}
\usepackage{cite}
\usepackage{siunitx}
\usepackage{comment}

\usepackage{xcolor}

\usepackage{hyperref}

\makeatletter
\newcommand{\printfnsymbol}[1]{%
  \textsuperscript{\@fnsymbol{#1}}}%

\begin{document}

\title{Temporal dynamics of GHz acoustic waves in chipscale phononic integrated circuits}

\author{A. Fahad Malik\IEEEauthorrefmark{2},
\and Mahmut Bicer\IEEEauthorrefmark{2},  \and Krishna C. Balram

\thanks{This work was supported in part by the UK's Engineering and Physical Sciences Research Council (EP/V005286/1, EP/N015126/1), the European Research Council (SBS 3-5, 758843) and the UKRI frontier research guarantee (EP/Z000688/1).}

\thanks{A.F. Malik gratefully acknowledges funding from the Commonwealth Scholarship Commission and the Foreign, Commonwealth and Development Office in the UK. All views expressed here are those of the author and not the funding body.}

\thanks{M. Bicer, A.F. Malik, and K.C. Balram are with the Quantum Engineering Technology Labs and the School of Electrical, Electronic and Mechanical Engineering, University of Bristol, Woodland Road, Bristol BS8 1UB, UK}
\thanks{\IEEEauthorrefmark{2} Contributed Equally}
\thanks{email: fahad.malik@bristol.ac.uk}
\thanks{email: mahmut.bicer@bristol.ac.uk}
\thanks{email: krishna.coimbatorebalram@bristol.ac.uk}}



\maketitle

\begin{abstract}
Phononic integrated circuits, which manipulate \qty{}{\GHz}-frequency acoustic fields in \qty{}{\um}-scale waveguides, provide new degrees of freedom for routing and manipulation of microwaves in deeply sub-wavelength geometries with associated implications for chipscale sensing and signal processing. The combination of low propagation loss, long interaction lengths and slow speed of sound put together with the large measurement bandwidths and high frequency resolution available from modern vector network analyzers (VNA) makes it feasible to visualize the temporal dynamics of propagating acoustic fields in these devices and \textit{see the device in action}. Two representative examples we discuss here are pulse circulation and ringdown in an acoustic microring resonator, and the observation of (parasitic) multipath interference effects in waveguide resonator geometries. In the absence of fast 3D acoustic field imaging modalities, such time domain reflectometry based methods provide the best alternative for mapping interface reflection and loss, which becomes increasingly critical as these devices start to scale in complexity. 
\end{abstract}


\section{Introduction}
\IEEEPARstart{P}iezoelectric phononic integrated circuits (PnICs) \cite{kcb2024piezoelectric}, platforms that can route \qty{}{GHz}-frequency acoustic fields in wavelength (\qty{}{\um}-scale) waveguide geometries, are currently being actively explored across a variety of material platforms \cite{fu2019phononic, mayor2021gigahertz, bicer2022gallium} for applications in sensing and signal processing. Like their well-established integrated photonics counterparts \cite{chrostowski2015silicon}, PnICs provide low-loss \cite{bicer2023low} waveguiding ($\approx$ \qty{3.6}{\decibel\per\mm} at \qty{3.4}{GHz}) with strong confinement and long interaction lengths in chipscale platforms. Given that the acoustic field is generated from a microwave signal in piezoelectric devices, PnIC platforms provide new degrees of freedom for manipulating microwaves in chipscale platforms at the deeply subwavelength-scale (acoustic wavelength ($\lambda_a$) $\approx$ \qty{}{\um}, RF wavelength $\approx$ \qty{}{\cm}). An illustrative example from our previous work \cite{bicer2023low} is the demonstration of compact spiral delay lines that achieve on-chip signal delays of $\approx$ \qty{2.5}{\micro\second}, corresponding to a free space delay of $\approx$ \qty{750}{\meter}, while maintaining an on-chip footprint $<$ \qty{0.25}{\milli\meter\squared}.

A large part of the promise of such chipscale acoustics platform, like with their microelectronic and photonic counterparts, derives from the prospect of cascading multiple elements together and getting the field (either electrons, light, or microwaves via sound) to flow through them in a controlled sequence to realize complex functionality. A simple example would be to go beyond the single acoustic microring resonator \cite{bicer2023low} towards coupled ring-resonator \cite{little1997microring} geometries with a view towards improving spectral filtering. But cascading elements brings with it the challenges of designing (multiple) interface transitions to minimize reflections and avoid mode conversion, as these parasitic reflections in a low-loss waveguide platform can lead to a variety of multipath interference effects, as we demonstrate below. These stray fields can cause spectral distortion, with the classic example being triplet transits in traditional surface acoustic wave (SAW) devices \cite{morgan2010surface}.

In the absence of (fast) 3D field imaging modalities that can map acoustic fields with nm-scale depth resolution \cite{zheng2020imaging}, time domain reflectometry (TDR) methods \cite{packard1998time} provide the best alternative for \textit{mapping} propagating acoustic fields in guided wave geometries. While TDR methods have long been used in the microwave domain \cite{bryant1993principles}, they are particularly suited to studying acoustic fields in PnICs. The combination of low propagation loss, long interaction lengths and slow speed of sound combined with the large measurement bandwidths and frequency resolution of modern VNAs makes it feasible to visualize the temporal dynamics of propagating \qty{}{GHz}-frequency acoustic fields in these devices in exquisite detail, as we discuss below in Section \ref{Section_3}. In a way, the PnIC platform provides a nice playground for acoustic dynamics, especially multipath interference effects, and the VNA has sufficient sensitivity for us to resolve and identify these events. This sensitivity is the main advantage of using the time domain transform of VNA spectral data over doing the experiments in the time domain \cite{bicer2023low}, as insertion loss \cite{kcb2024piezoelectric} is currently the main bottleneck in these chipscale devices and limits the achievable signal-to-noise ratio severely in practice.

More broadly, visualizing the temporal dynamics allows us to \textit{see the device in action}. Although PnICs have been developed in analogy with their photonic counterparts, the slower speed of sound coupled with direct access to the phase information in a VNA makes studying temporal dynamics in this platform much more accessible and provides additional insights that cannot be directly inferred from the spectral domain. Here, we use the side-coupled waveguide ring resonator geometry \cite{bicer2022gallium, bicer2023low} as a model device to illustrate these ideas.

\section{Acoustic Microring resonators} \label{Section_2}

\begin{figure}[!htbp]
    \includegraphics[width =0.9 \columnwidth]{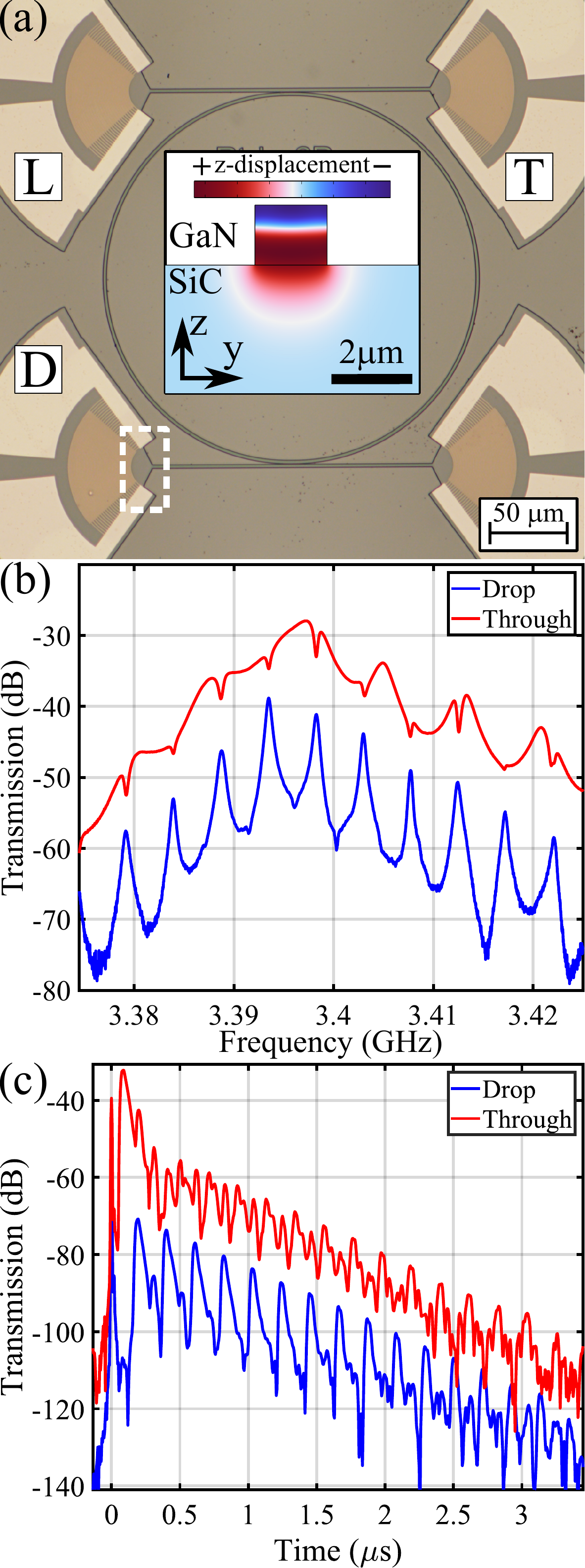}
    \caption{(a) Microscope image of a representative acoustic microring resonator device. Inset shows an FEM simulation of the $z$-displacement of the propagating waveguide mode. Focusing interdigitated transducers (IDTs) are used to convert microwave fields at $\approx$ \qty{3.4}{GHz} into propagating acoustic fields (launch port, L) which can be efficiently routed using on-chip waveguides. The acoustic field in the waveguide can couple into and circulate in an acoustic microring resonator. The power leaking out from the resonator can be measured in both the through (T) and drop (D) ports. (b) The transmission spectrum of the device in the T (red) and D (blue) ports. (c) Time domain transform of the data in (b) for the T (red) and D (blue) ports. The time-domain T port data is shifted by \qty{20}{\decibel} for illustration clarity.}
    \label{Fig1_AMR_FT}
\end{figure}

Fig. \ref{Fig1_AMR_FT}(a) shows an optical microscope image of a representative acoustic microring resonator device fabricated in a gallium nitride (GaN) on silicon carbide (SiC) platform \cite{bicer2023low}, with radius ($R$) of \qty{115}{\um} side-coupled to acoustic waveguides with width ($w_{wg}$) of \qty{1.8}{\um} and thickness ($t_{wg})$ of \qty{1.5}{\um}. The waveguide resonator gap ($g$) is $\approx$ \qty{600}{\nm}. Focusing interdigitated transducers (FIDTs) are used to convert microwave fields at $\approx$ \qty{3.4}{GHz} into propagating acoustic fields that are focused at the waveguide entrance for efficient mode matching. A finite element method (FEM) simulation of the waveguide mode is shown in the inset of Fig.\ref{Fig1_AMR_FT}(a). At the output port of the waveguides, the process is reversed and the FIDTs are used to reconvert the acoustic field exiting the waveguide into the microwave domain. The microwave signal is generated and detected in a standard swept-frequency scattering matrix (S-parameter) measurement using a vector network analyzer (Keysight E5063A). 

The resonator transmission spectrum ($S_{21}$) measured in the through (T, red) and drop (D, blue) ports are shown in Fig.\ref{Fig1_AMR_FT}(b). This device has a coupling length ($l_{cp}$) of \qty{6}{\lambda_a} where the acoustic wavelength ($\lambda_a$), which determines the IDT period,  is \qty{1.6}{\um}. The waveguide resonator gap in this case is \qty{500}{\nm}. Both transmission spectra show a periodic series of resonance peaks in D (dips in the T), that correspond to successive resonant modes of the ring resonator. The resonance condition can be stated approximately as $m\lambda_{R}=2{\pi}R$ where $m$ is an integer, $\lambda_R$ is the acoustic wavelength in the ring and $R$ is the ring radius. Each of the successive resonances correspond to successive integer values of $m$. As noted previously \cite{bicer2023low}, acoustic ring resonators relying on total internal reflection of acoustic waves can have very low intrinsic damping, approaching the underlying material limits, and can exceed the loss performance of traditional SAW and bulk acoustic wave (BAW) \cite{hashimoto2009rf} based resonators at the same frequency. 

We can numerically transform the complex spectral data into the time domain \cite{wollensack2012vna}. The signal transmission as a function of time is shown in Fig.\ref{Fig1_AMR_FT}(c) with the T and D ports indicated in red and blue respectively, and the T port data has been offset by \qty{20}{\decibel} for clarity. We would like to note that the T and D port measurements are not done simultaneously because we are limited to a 2-port VNA. The ringdown of the resonator is clearly seen in both the T and D port data, with each successive temporal pulse corresponding to one additional round trip of the cavity. The difference in arrival time of subsequent pulses allows us to estimate the group velocity ($v_g$) to be \qty[separate-uncertainty = true,multi-part-units=single]{3534\pm25}{\meter\per\second} which agrees well with the frequency domain estimate of \qty[separate-uncertainty = true,multi-part-units=single]{3526\pm33}{\meter\per\second} calculated from the measured cavity free spectral range (FSR $\approx$ \qty[separate-uncertainty = true,multi-part-units=single]{4.754\pm0.045}{\MHz}). The difference is mainly due to the finite temporal resolution (\qty{5}{\nano\second}) in the system which bounds the fidelity with which we can estimate the arrival time of each pulse. This time resolution is given by: 

\begin{equation}
dt = \frac{1}{n\cdot df}\ \raisebox{0.5ex}{,}
\end{equation} where $n$ is the number of frequency points (usually \qty{1e4}{}) and $df$ is the frequency spacing ($\approx$ \qty{20}{\kHz})\cite{wollensack2020metas}. The sensitivity and high dynamic range afforded by the VNA is exemplified by the fact that we observe up to 15 pulses before the signal gets buried in the noise. We would like to emphasize here that the temporal pulses correspond (taken as a weighted sum) to all the resonant modes of the cavity that lie within the measurement frequency span, and the effective loss from fitting the ringdown should be interpreted within this weighted average limit, and cannot be ascribed to any individual resonance. We extract a $1/e$ (power) ringdown time of $\approx$ \qty{280}{\nano\second} which agrees well with the Lorentzian fitting of the individual cavity resonances in the frequency domain, giving us modal quality factors \cite{petersan1998measurement} in the range of $\approx$ \qty{6000}{}, consistent with previous results \cite{bicer2023low}.

\begin{figure}[htbp]
    \includegraphics[width =0.95 \columnwidth]{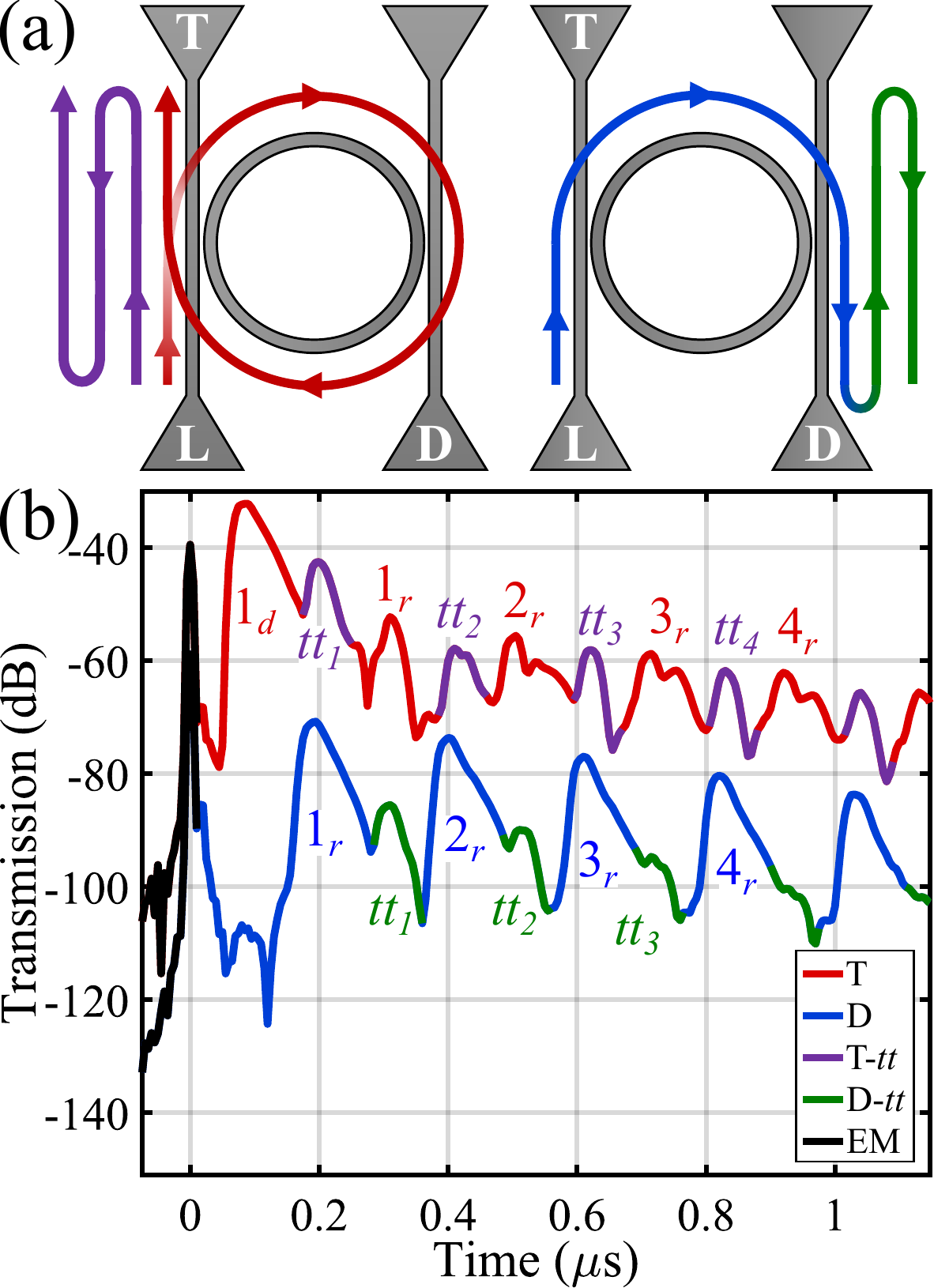}
    \caption{(a) Schematic of the possible paths taken by the acoustic field from transmit to receive ports (either through or drop). The reflection at the waveguide taper interface, coupled with the low propagation loss leads to interference signals in the receive port that do not correspond to ring transits. (b) Zoomed-in time domain transmission response of the device from Fig.\ref{Fig1_AMR_FT}(c) for both the drop (blue) and through (red). The through port data has been shifted by \qty{20}{\decibel} for clarity. The successive ring transits are labelled by $m_r$ and the triplet transits by $tt_i$, where $m$ and $i$ are integers. The residual EM crosstalk is shown in black.}
    \label{Fig2_zoomin}
\end{figure}

If we zoom-in to the time domain dataset in Fig.\ref{Fig1_AMR_FT}(c), replotted in Fig.\ref{Fig2_zoomin}(b) for clarity, we can get further insight into the temporal dynamics. Fig.\ref{Fig2_zoomin}(b) clearly shows the residual electromagnetic (EM) crosstalk (black) between the transmit (L) and receive (T,D) ports arriving almost instantaneously followed by the train of acoustic pulses. In the T port, the first pulse (indicated $1_{d}$) corresponds to the direct transit that skips the ring resonator whereas the subsequent pulses that couple into and out of the ring resonator are labelled by $m_{r}$, where $m$ is the number of ring transits a pulse undergoes before detection. We can unambiguously identify the ring transits by noting that the D and T ports should have pulses that differ by $\approx{\pi}R/v_g$ as the drop pulse leaks out at the mid-way point in the ring. The schematic in Fig.\ref{Fig2_zoomin}(a) illustrates the paths and the data in Fig.\ref{Fig2_zoomin}(b) is color coded to match the paths. We label the D port peaks also by the $m_{r}$ notation to show the correspondence between the two ports and emphasize that it is the same circulating pulse that leaked out at different instances which leads to the difference in detection times. 

We can also observe an additional set of peaks, shaded purple and green respectively, in the T and D port time domain response in Fig.\ref{Fig2_zoomin}, labelled $tt_{i}$, where $i$ is an integer. These are analogous to triplet transits in SAW devices \cite{morgan2010surface}. The paths taken by the first triplet transits in the T and D ports are indicated in the schematic in Fig.\ref{Fig2_zoomin}. Each of these secondary pulses that follows the cavity pulse occurs due to the reflection of the cavity pulse at the waveguide taper interface, cf. dashed white box in Fig.\ref{Fig1_AMR_FT}(a). We can verify this is the case by noting that the waveguide length is \qty{176.6}{\mu m} and given a $v_g$ of \qty{3534}{\meter\per\second}, these $tt$ pulses should experience an additional delay of \qty{100}{\nano\second} and we measure a delay of $\approx$ \qty{110}{\nano\second}. In the T port data, one can also see that the pulse shape is distorted (cf. $m\geq$ 2), and one can clearly see a second pulse arriving within the same time window. This is due to multi-path interference effects which we will discuss in detail in Section \ref{Mpath_int} below.

\section{Seeing the device in action: building up the resonance} \label{Section_3}

\begin{figure}[htbp]
    \includegraphics[width =0.95 \columnwidth]{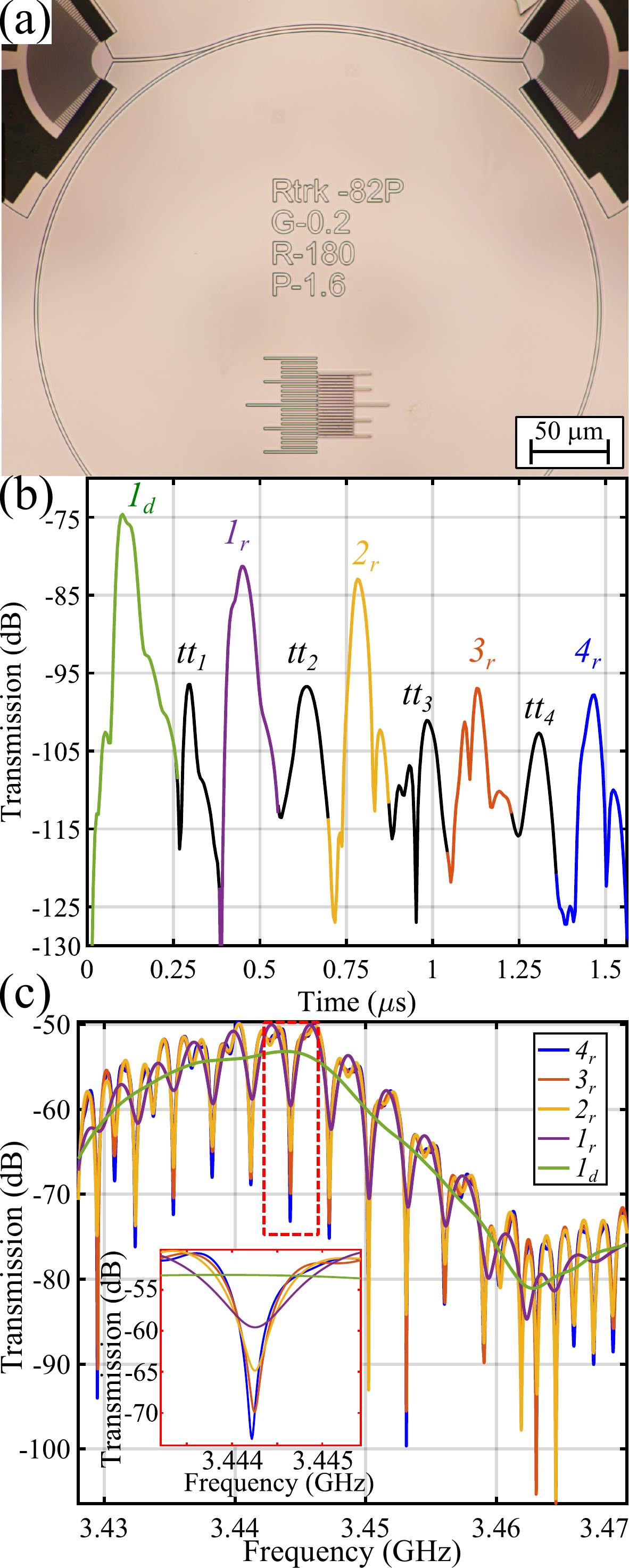}
    \caption{(a) A representative optical image of a pulley coupled microring resonator with ring radius \qty{180}{\micro\meter}, gap ($g$) \qty{200}{\nano\meter}, and coupling length ($l_{cp}$) \qty{80}{\lambda}. (b) A time-domain representation of the transmission data of a pulley coupled microring device with $l_{cp}$ of \qty{102}{\lambda}. The labelling of the pulses follows the nomenclature of Fig.\ref{Fig2_zoomin}. (c) The transmission spectrum of the pulley coupled ring, showing the build-up of the resonance as subsequent ring transits are accounted for. The legend (and colour scheme) correspond to the peaks indicated in (b) and indicates the width of the temporal bandpass filter: $i_r$ means the time domain trace from $t=0$ until the end of $i_r$ is taken to generate the corresponding spectral response.}
    \label{Fig3_res_buildup}
\end{figure}

As noted in the introduction, the combination of bandwidth, frequency resolution and dynamic range afforded by modern VNAs with the low loss, long interaction lengths and slow speed of sound in PnICs makes it possible to directly visualize the device operation. We again use the microring resonator to illustrate this idea because the device operation is well understood analytically \cite{bogaerts2012silicon}, but this methodology can be extended to more complicated instances with multiple interacting components (and delayed feedback loops), wherein analytical solutions might not be tractable. 

Fig.\ref{Fig3_res_buildup}(a) shows a micrograph of a pulley-coupled microring resonator. The pulley-coupled design increases the effective waveguide resonator coupling length and pushes the device operation closer to the critical coupling regime wherein the cavity decay rate is exactly matched by the waveguide resonator coupling rate and the device transmission is exactly zero at resonance. The point coupled ring in Fig.\ref{Fig1_AMR_FT}(a) instead operates in an undercoupled regime wherein the cavity decay rate exceeds the waveguide coupling rate. The larger extinction depth in transmission makes device visualization easier, although the principle of operation of the two devices are (nearly) identical.

Fig.\ref{Fig3_res_buildup}(b) shows the temporal transmission response of the device with the peaks labelled according to the notation in Fig.\ref{Fig2_zoomin}: $m_r$ successive ring transits, $tt_i$ associated triplet transit and the first pulse ($1_d$) corresponding to a direct waveguide transit (skipping the ring). We have removed the EM crosstalk by applying a Hann-shaped notch filter (center \qty{0}{\nano\second}, span \qty{100}{\nano\second}) \cite{wollensack2020metas} to the time domain data. Fig.\ref{Fig3_res_buildup}(c) shows the change in the spectral transmission response of the device as each subsequent ring transit pulse is accounted for. We perform this operation by applying a bandpass filter to the time domain data, centred at $t=$ \qty{0}{}, with increasing filter width to accommodate subsequent pulses, before transforming the data back to the frequency domain. We note that the bandpass filtering includes the triplet transit pulses in addition to the ring transits. A zoom-in to one of the resonances is shown in the inset of Fig. \ref{Fig3_res_buildup}(c) which clearly shows the evolution of the resonance lineshape from a broad background (green curve, corresponding to the direct transit) to a sharp Lorentzian lineshape whose spectral width reduces and extinction depth increases as more pulses (each delayed by one cavity round trip) are combined together by increasing the temporal bandpass filter width. While this result is well understood from Fourier analysis, being able to empirically measure this result, and explicitly verify the device operation is illuminating.

As Fig.\ref{Fig3_res_buildup}(c) (and the inset, which zooms into one of the modes) clearly show, the (subsequent) pulses from the ring resonator arrive in phase (modulo 2$\pi$) and are always $\approx\pi$ out of phase with the direct transmission (green curve). The $\pi$ phase difference is usually understood by treating the waveguide resonator coupling region as a partially transmitting beam splitter \cite{yariv2002critical} and using unitarity to impose a ${\pi}/2$ phase shift between the reflected and transmitted beams referenced to two suitably chosen reference planes. Given the ring channel encounters the coupler twice while coupling in and out, the (overall) phase difference with respect to the waveguide channel is $\pi$. While this result is well understood analytically \cite{yariv2002critical, bogaerts2012silicon, chrostowski2015silicon}, it is still notable that it holds in a real device like in Fig.\ref{Fig3_res_buildup}(a), which does not fit the blackbox model of a lossless beam splitter with point coupling \cite{hamilton2000phase}, wherein such general symmetry considerations can be invoked. More precisely, the observation of multi-cycle interference (between subsequent round trips of the cavity) shows that insertion loss and mode conversion are negligible even in a long distributed pulley coupling interface like Fig.\ref{Fig3_res_buildup}(a). Moreover, we can also conclude that (residual) surface roughness in the waveguide does not induce appreciable mode conversion, and effectively single-mode propagation and interference can be observed in an intrinsically multimode system, with the caveat that the IDTs are in practice mode-selective receivers, and we can not distinguish between mode conversion and excess attenuation in these experiments. 

\begin{figure}[htbp]
    \includegraphics[width =0.95 \columnwidth]{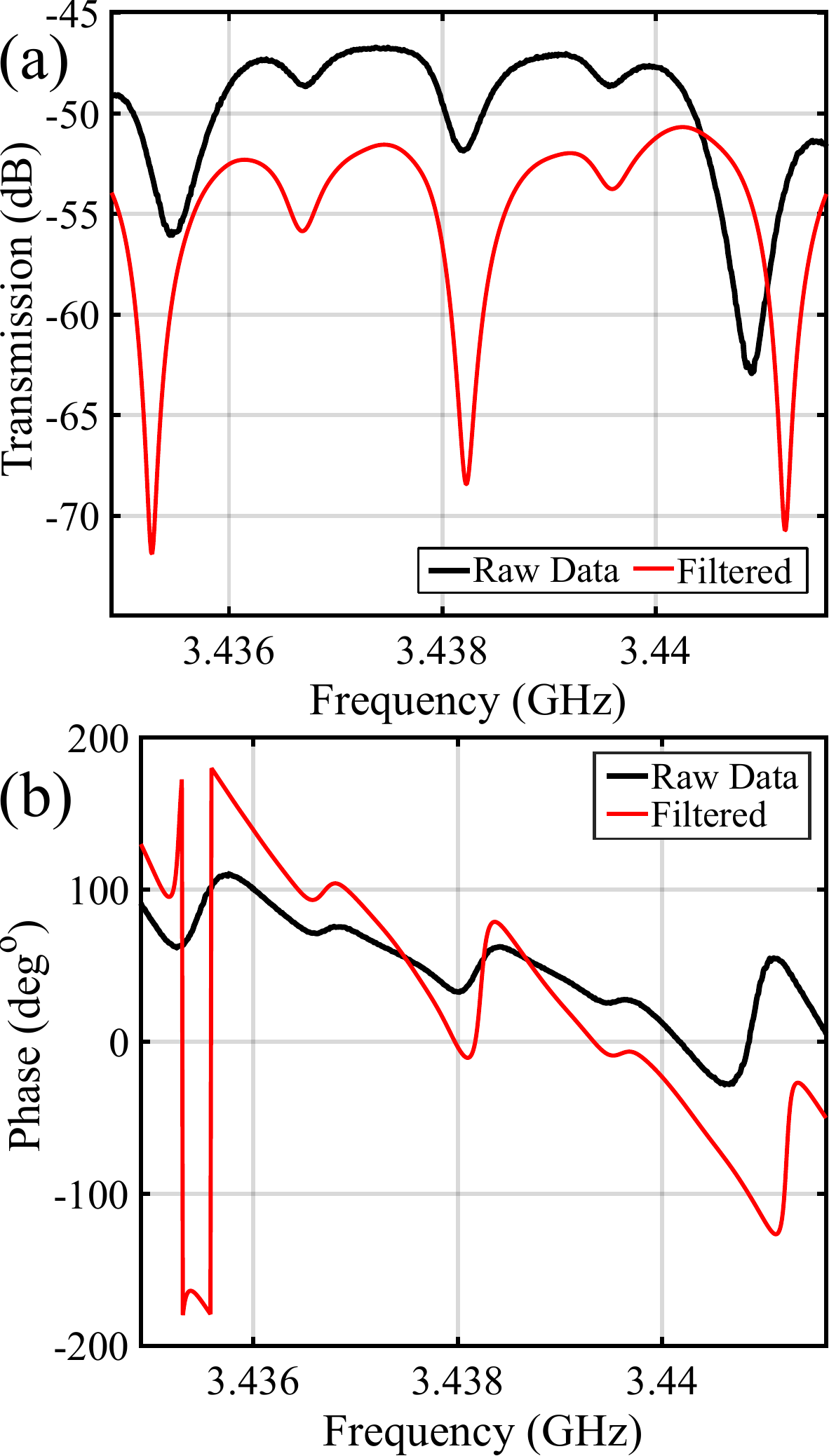}
    \caption{(a) Zoomed-in transmission spectrum (magnitude) of the device from Fig.\ref{Fig3_res_buildup} showing the effect of filtering out the EM crosstalk and the triplet transits (red curve) on the raw data (black) (b) Effect of filtering on the phase response. In this device, the EM crosstalk has a far greater effect than the $tt$, but as we move towards impedance matched transducers and more complicated circuit elements, the situation is expected to be reversed.}
    \label{Fig4_parasitic_filtering}
\end{figure}

In addition to providing a graphic visualization of how the cavity resonance builds up, the ability to selectively isolate features in the time domain data allows us to infer the spectral distortion induced by unwanted (parasitic) processes. PnIC platforms provide a natural route towards spectral shaping and control of microwave transmission by cascading multiple resonant elements together in arbitrary geometries. However, the desired spectral response can only be achieved if the different interfaces are close to lossless, i.e. the different elements can be concatenated without significant signal back-reflection. The effect of this spectral distortion on the ring resonator transfer function can be seen in Fig.\ref{Fig4_parasitic_filtering}, which shows a zoomed-in spectral transmission of the pulley coupled device from Fig.\ref{Fig3_res_buildup}(a). The as-measured (raw data) spectral transmission is shown in black (Fig.\ref{Fig4_parasitic_filtering}). By applying a notch filter of \qty{100}{\nano\second} width \cite{wollensack2020metas}, centred at $t=0$, we can remove the EM crosstalk. By further applying a series of notch filters (time gates) of \qty{200}{\nano\second} width centred on the triple-transit peaks (labelled $tt$ in Fig.\ref{Fig3_res_buildup}(b)), we can remove all the triplet transits from the data in the time domain before transforming back to the frequency domain. This filtered data, with the EM crosstalk and triple transits removed, is shown by the red curve labelled filtered in Fig. \ref{Fig4_parasitic_filtering}. One can clearly see that both the magnitude and the phase response are dramatically improved once the EM crosstalk is removed. Removing the triplet transits has a negligible effect in this case, but will become increasingly important in more complicated devices envisioned in future PnIC platforms. One feature of note is that the EM filtering has changed the spectral position of the resonator modes, seen most easily in Fig.\ref{Fig4_parasitic_filtering}(a) by noting the location of the dips for the black and red curves. We believe this occurs because the EM background has a non-trivial phase profile, but we don't quantitatively understand the shift, especially as the shift in frequency appears random, cf. the redshift for the first dip, negligible shift for the second and a blue shift for the third dip in Fig.\ref{Fig4_parasitic_filtering}(a). 

\section{Quantifying multipath interference} \label{Mpath_int}

While the discussion so far has focused on elucidating the temporal dynamics of \qty{}{\GHz} acoustic fields in PnICs by identifying the different features observable in the temporal transform of the VNA data, here we show that we can quantitatively model, with some simplifications, the scattering response of the device and extract the interface reflection coefficients that cause the multipath interference effects. As noted above, one of the main challenges \cite{kcb2024piezoelectric} in scaling PnIC circuits from $O(1)$ to $O(10)$ spectral components in the near-term is minimizing these interface reflections.

\begin{figure}[htbp]
    \includegraphics[width =1 \columnwidth]{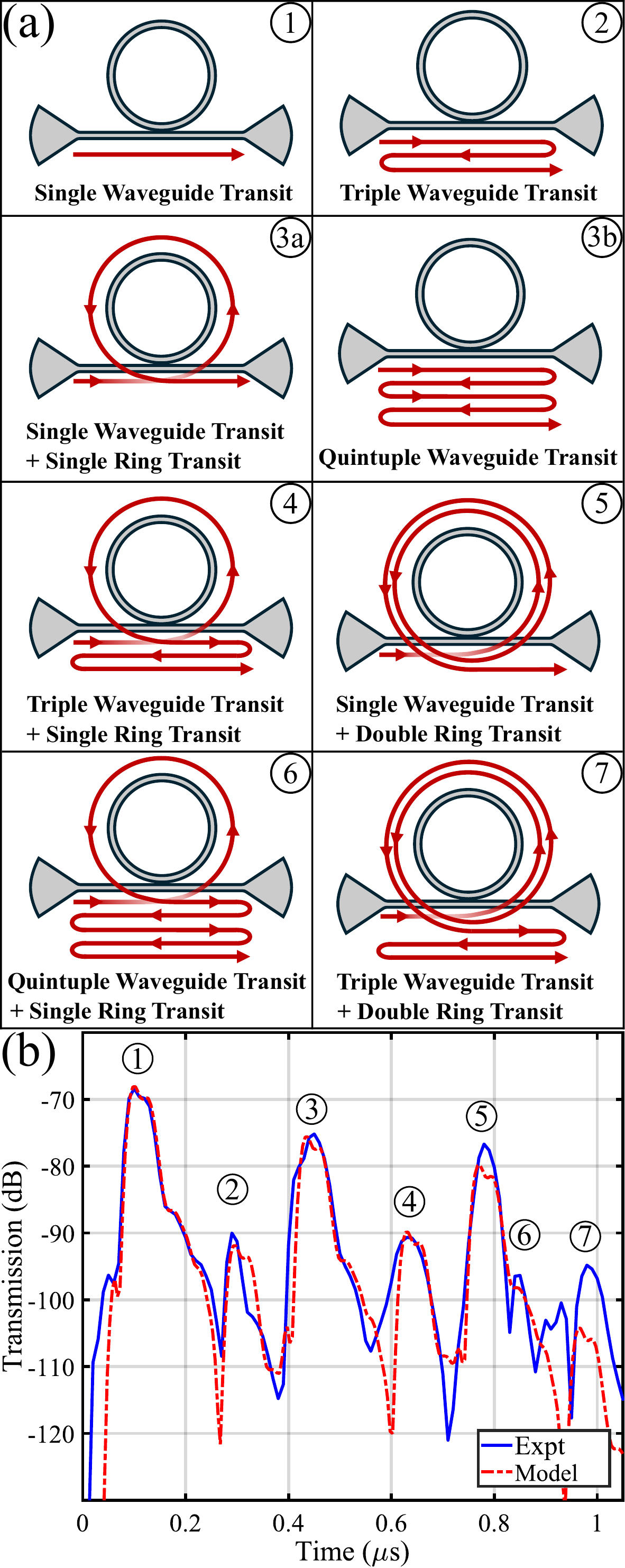}
    \caption{(a) Schematic representation of the various possible paths the acoustic signal can take from transmitter to receiver, arranged in chronological order, grouped by temporal delay (b) Temporal transmission (blue line) of the pulley-coupled ring device, along with a best-fit model prediction (red, cf. Appendix \ref{App:model} for details).}
    \label{Fig5_graph}
\end{figure}

We return to the pulley coupled device shown in Fig.\ref{Fig3_res_buildup}(a) and re-plot the time domain transmission in Fig.\ref{Fig5_graph}(b). In Fig.\ref{Fig5_graph}(a), we graphically outline the different potential routes (numbered chronologically) the signal can take from transmitter to receiver and indicate the corresponding temporal delay for each path in Fig.\ref{Fig5_graph}(b). It is important to keep in mind that the waveguide resonator junction functions as an acoustic beam splitter. Therefore, for instance the path labelled 3(a) in Fig.\ref{Fig5_graph}(a), single waveguide transit and single ring transit, is actually traversed in the following order: the pulse propagates half the waveguide length, couples into the ring, completes a ring transit and couples back out into the waveguide wherein it traverses the second half of the waveguide. To illustrate another example, path 5 single waveguide transit and double ring transit starts out identical to path 3(a), but after the first ring transit, we are here tracking the pulse component that stays within the ring when it encounters the beam splitter, undergoes another transit of the ring and the couples out into the waveguide. The diagrams in Fig.\ref{Fig5_graph}(a), especially those with ring transits involved should be interpreted in this schema.

We can construct a simplified model of the system response (cf. Appendix \ref{App:model} for further details) by noting, as a first approximation, that the complete time response trace is made by the same pulse shape taking different paths (with corresponding delays) from transmitter to receiver. This simplification allows us to use the measured first arriving pulse (the direct transit) as a template to generate the complete time response by adding the appropriate delay for each paths and the corresponding attenuation to fit the measured response (cf. Appendix \ref{App:model} for details on the fitting procedure). Using the measured pulse response allows us to work around the problem of accounting for the spectral distortion introduced by the (transmit and receive) IDTs and the complex phase shifts at the waveguide taper interface. Measuring the excess attenuation between a ring transit and its associated triplet transit (for ex: between paths 3a and 4 in Fig.\ref{Fig5_graph}(a)) and noting that the triplet transit undergoes two additional reflections at the waveguide taper interface in addition to the (known) propagation loss encountered in traversing the straight waveguide section back and forth, we can extract the interface (power) reflection coefficient ($\mathcal{R}$). Given the waveguide taper interface (dashed white box in Fig. \ref{Fig1_AMR_FT}(a)) stays the same, regardless of the waveguide-resonator geometry, we can apply this procedure to extract $\mathcal{R}$ across all devices to improve the estimate. We extract an overall $\mathcal{R}\approx$ \qty{0.21}{}$\pm$0.07, but find in practice, that there is a significant device-to-device variation beyond what we can attribute to fabrication imperfections, cf. Appendix \ref{App:model} for representative datasets. We would also like to clarify that, given the temporal resolution of our system, we can only get a net reflection coefficient and not the individual components due to mode-mismatch at the waveguide taper interface, and the reflection from the IDT fingers (Cr/Au) due to (mechanical) impedance mismatch. 

\begin{figure}[htbp]
    \includegraphics[width =1 \columnwidth]{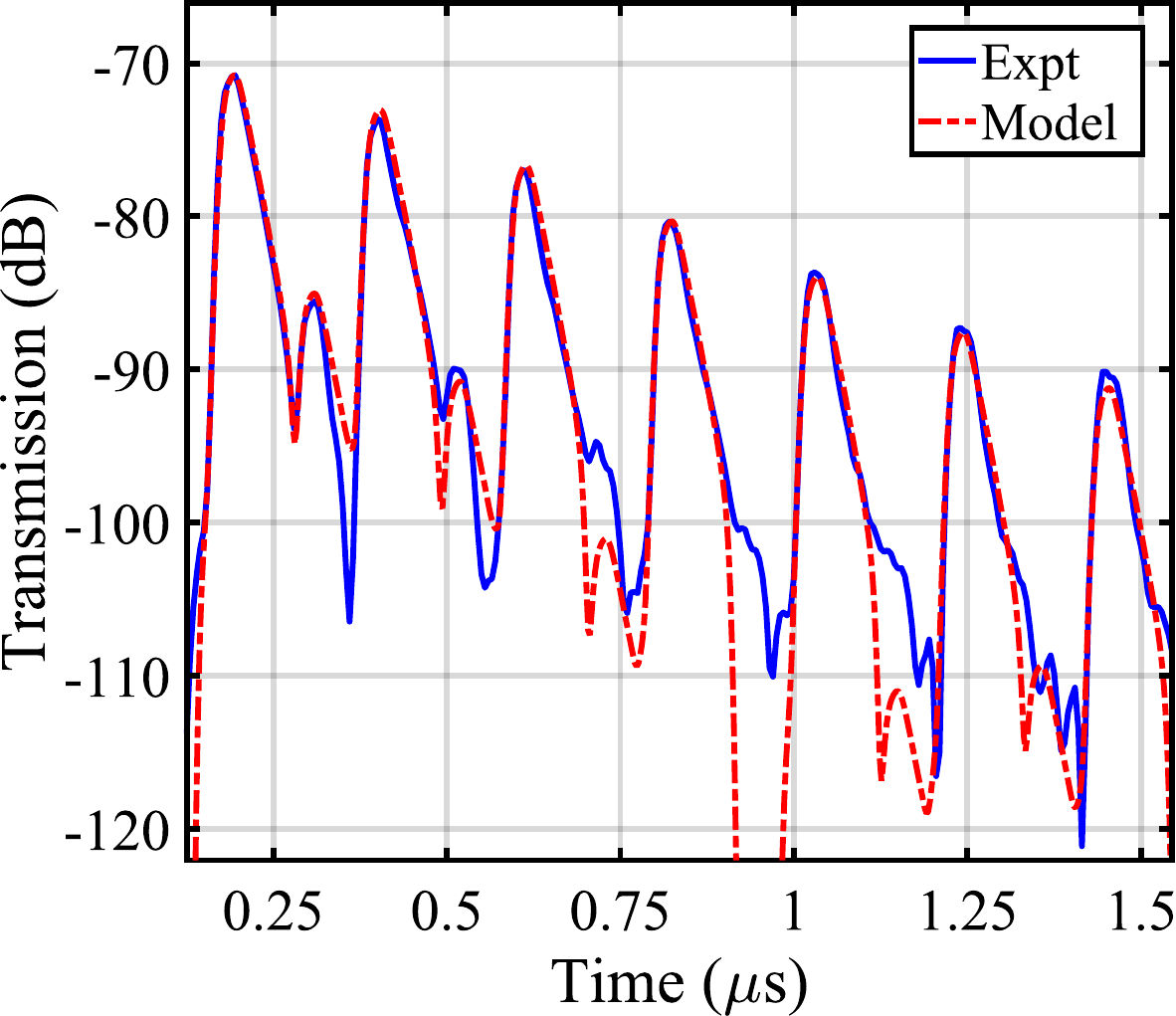}
    \caption{ Drop port time domain transmission (blue), and predictive model fit (red, cf. Appendix\ref{App:model}) for a four port ring resonator device, as shown in Fig.\ref{Fig1_AMR_FT}(a). The drop port data is cleaner and fits better than the through port (cf. Fig.\ref{Fig5_graph}(b) due to the spectral and modal filtering induced by the ring and the avoidance of the background straight through path ($1_d$ in Fig.\ref{Fig2_zoomin}(b),\ref{Fig3_res_buildup}(b))}
    \label{Fig6_graph}
\end{figure}

By fitting the overall temporal response using this procedure as shown in Fig.\ref{Fig5_graph}(b) and knowing the different trajectories (with their different relative losses), we can extract the full waveguide resonator coupling matrix \cite{bogaerts2012silicon}, cf. Table I in Appendix \ref{App:model}. As expected from using the direct transit pulse shape as a template, we get better agreement with predicting the pulse arrival time, but our simplified model doesn't fully capture all the pulse distortion effects leading to differences between the predicted pulse shapes and the experimental measurements. We would like to note that some of the sources of error are down to the intrinsic temporal resolution of our experiments that makes some of these multipath events hard to resolve (cf. events 5 and 6 from Fig.\ref{Fig5_graph}(a)). A second source of error in our model prediction is that we are simplifying some of the interactions and not fully accounting for the complex phase. For instance, the waveguide resonator coupling, even for the pulley coupled device in Fig.\ref{Fig3_res_buildup}(a) is modelled using an effective point coupling matrix \cite{yariv2002critical}. These errors are not detrimental in the current instance and can be used effectively for parameter extraction, as evidenced by the better model fit to the drop port data shown in Fig.\ref{Fig6_graph} for a 4-port device. Fig.\ref{Fig6_graph} clearly shows the spectral and modal filtering effects of the ring resonator as the fitting errors for the drop port data (using this template model) are significantly lower than in the through port.  

As discussed above, interface reflection in a low-loss waveguide platform is the main source of spectral (phase) distortion, and while these time domain measurements can uniquely pinpoint the different interference terms, it is an open question how well they can be mitigated \cite{kcb2024piezoelectric}. It is important to note here that this interface reflection ($\mathcal{R}\approx$ \qty{0.21}{}$\pm$0.07) issue for PnICs is significantly worse than for analogous photonic circuits ($\mathcal{R}\ll$ \qty{0.01}{}). This is due to a combination of reasons outlined in \cite{kcb2024piezoelectric}: an intrinsically multi-modal system with slow surface modes and the need to simultaneously satisfy electrical impedance matching and acoustic mode matching constraints \cite{korovin2019conversion}. Tackling this interface problem and ensuring one can get \qty{}{\GHz} acoustic fields into and out of \qty{}{\um}-scale devices with near-unity efficiency is critical to ensuring that PnIC platforms can scale to $O(10)$ components in the near-term.

\section{Conclusions}

In this work, we have shown that the temporal dynamics of propagating \qty{}{\GHz} acoustic fields in \qty{}{\um}-scale waveguide geometries in PnIC platforms can be accurately visualized and modelled using the time domain transform of the spectral data from a vector network analyzer. The combination of low propagation loss, long interaction lengths, slow speed of sound and multiple interfaces with the high dynamic range and sensitivity that the VNA data provides enables us to directly \textit{see the device in action}. We use an acoustic microring resonator geometry, excited by an on-chip waveguide, as a model platform and demonstrate resonant pulse circulation and ringdown and the quantification of multipath interference effects that occur due to reflection at component interfaces. As PnIC platforms scale in complexity, such temporal field mapping techniques become increasingly critical as both visualization aids and metrology tools for measuring and improving performance. 

\section{Acknowledgments}

We would like to thank Martin Cryan, John Haine, Martin Kuball, Bruce Drinkwater, Jacob Brown, Yingqiang He, Robert Thomas, Chaowang Liu, Hugues Marchand, and Yutian Zhang for valuable suggestions and feedback.

\appendices

\section{Modelling multipath interference effects} \label{App:model}
In order to capture the temporal propagation and interference effects observed in a ring resonator, we employ a multi-path interference model. In addition to the principal paths from the launch port to the drop/through port with or without ring traversals, we also consider additional subsidiary paths that arise when reflections from waveguide taper junctions are non-negligible. Thus, the signal from the transmit (launch) port can traverse a range of possible routes, encompassing direct paths, ring transits, and multiple triplet transits, some of which are outlined in Fig.\ref{Fig5_graph}(a), before emerging at the output port. Each of these routes contributes a distinct amplitude and phase to the total through/drop-signal as a function of time. Interference effects can be observed when the arrival time of two distinct paths lie within the temporal pulse width.

\begin{figure}[htbp]
    \includegraphics[width =1 \columnwidth]{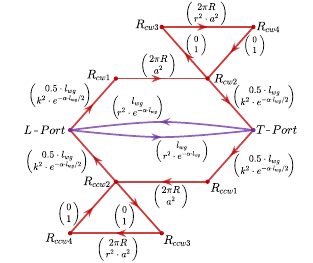}
\caption{Directed graph representation of a 2-port ring resonator used to model valid signal paths. Each edge is annotated with a 2×1 array indicating the physical propagation length and corresponding power transmission. The graph captures all possible routes the signal can take from the launch (L) port to the through (T) port, including multiple ring transits and various triplet transits due to reflection at the waveguide taper interface. Both clockwise (CW) and counterclockwise (CCW) roundtrips (cf. Fig.\ref{FigA2_paths}) within the ring are represented using four nodes each to fully capture the propagation behaviour. This allows modelling of cumulative coupling loss, with each successive roundtrip incurring an additional loss corresponding to $k^2$. Additional node losses at the ports due to reflection ($\mathcal{R}$) are omitted from the graph for clarity. This graph forms the basis for modelling multipath interference effects in the time-domain response.}
    \label{FigA1_graph}
\end{figure}

To enumerate all such paths, we represent the ring resonator as a directed graph as in Fig.\ref{FigA1_graph}, whose nodes correspond to physical locations or geometrical features. Directed edges connecting these nodes have weights corresponding to the propagation distance ($l_{prop}$), and associated loss due to propagation and coupling. Nodes corresponding to ports are also assigned weights corresponding to the (lumped) loss due to reflection at waveguide taper interface and IDTs. A path is then defined as a sequence of connected edges leading from the input node to the output node. A recursive pathfinding algorithm is employed to find valid paths that the signal can take in the directed graph. The algorithm is tweaked to allow for revisiting nodes after reaching the output node to account for reflections from the waveguide taper interface. Since this directed graph represents an Infinite Impulse Response (IIR) system, the number of valid paths grows exponentially with time. Therefore, we impose a maximum time/distance constraint based on the length of the time data.

There are 6 independent fit parameters in the model (5 for 2-port ring resonator). These parameters, following the nomenclature of \cite{bogaerts2012silicon}, are: (i) the ring single-pass amplitude transmission coefficient $a$ related to the attenuation coefficient $\alpha$ [\qty{}{\per\cm}] as $a^2=e^{-{\alpha}2{\pi}R}$ with $R$ the ring radius; (ii, iii) the self-coupling coefficients at the two waveguide resonator interfaces for an add-drop device, $r_1$ and $r_2$. For the point coupled ring resonators, by symmetry $r_1=r_2$ and we also assume the coupling is lossless, hence $k_i^2+r_i^2=1$ with $k$ the cross-coupling (waveguide to ring) coefficient; (iv) an \textit{effective} increase in waveguide length, $l_{eff} = l_{wg}+\delta l$, to account for the (slight) difference in $v_g$ in straight waveguide and the curved ring sections; (v) \textit{effective} length of IDT region $l_{idt}$; and finally (vi) the interface (power) reflection coefficient $\mathcal{R}$. These parameters are determined using the algorithm below:

\begin{algorithm}[H]
\caption{\textbf{Parameter Estimation from Time-Domain Response}}
\textbf{Input:} Ring Radius $R$, Waveguide Length $l_{wg}$, Transmission Spectrum  \\
\textbf{Output:}{ $a$, $r_1$, $r_2$, $l_{\text{eff}}$, $l_{\text{idt}}$, $\mathcal{R}$}
\begin{algorithmic}[1]
\State Calculate $v_g$ from the measured cavity FSR
\State Fit the magnitude of peaks $\mathit{1}_d$ and $n_r$ to estimate initial values of $a$, $r_1$, $r_2$
\While{$\mathit{1}_d$ and $tt_1$ are not aligned in time}
    \State Increase $\delta l$
    \State Update $l_{\text{eff}} = l_{\text{wg}} + \delta l$
\EndWhile
\While {model and measured peaks are not aligned in time}
    \State Increase $l_{idt}$
\EndWhile
\State Estimate $\mathcal{R}$ by minimizing:
\[
\min_{\mathcal{R} \ge 0} \; \mathcal{F}(\mathcal{R})
\]
\State \Return $a$, $r_1$, $r_2$, $l_{\text{eff}}$, $l_{\text{idt}}$, $\mathcal{R}$
\end{algorithmic}
\end{algorithm}

As discussed in the main text, the interface reflection coefficient ($\mathcal{R}$) can be extracted by comparing the ratio between the received signal power of a ring transit and its associated triplet transit. For drop port $\mathcal{R}$, this is done numerically by minimizing a function $\mathcal{F}(\mathcal{R})$, subject to constraint $\mathcal{R} \ge 0$, defined as follows for a representative case:

\begin{equation}
\min_{\mathcal{R} \geq 0} \mathcal{F}(\mathcal{R}) 
= \min_{\mathcal{R} \geq 0} 
\Biggl| 
\frac{\hat{P_{tt_1}}}{\hat{P_{1_r}}} - 
\left\lvert \ell_1 + \ell_2 + \ell_3 \right\rvert^2 \\[1em]
\Biggr|,
\label{eq:interfaceReflection_opt}
\end{equation} 

\begin{figure}[htbp]
    \includegraphics[width =1 \columnwidth]{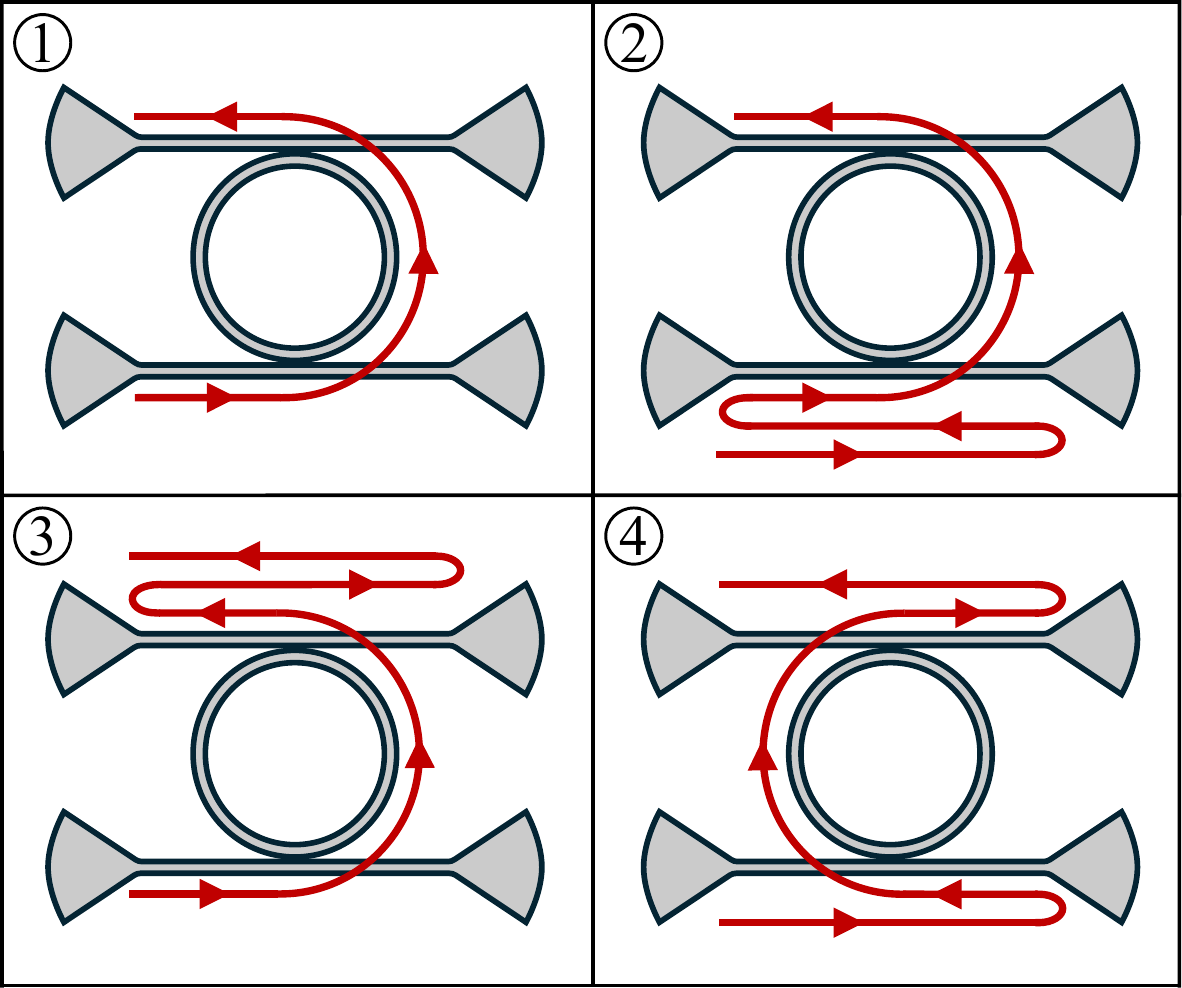}
    \caption{Schematic representation of the paths used to estimate $\mathcal{R}$ by taking the power ratio between a ring transit and its associated triplet transit. Path 1 is the first drop-port ring traversal and corresponds to Peak $\mathit{1_r}$. Paths labelled 2--4 correspond to the three permutations of the signal path comprising triple waveguide and half ring transit. Signals from Paths 2-4 arrive in-phase but may have different signal strength (if $r_1 \ne r_2$). Note that the parasitic paths involve both clockwise and counterclockwise ring transits on the way from transmitter to receiver IDT.}
    \label{FigA2_paths}
\end{figure}

Here, we extract $\mathcal{R}$ by comparing the signal amplitude of a direct ring transit and the associated triplet transits, shown in Fig.\ref{FigA2_paths}. $\hat{P_{tt_1}}$ and $\hat{P_{1_r}}$ denote the measured peak powers of $tt_1$ and $1_r$ (cf. Fig.\ref{Fig2_zoomin}(b)), respectively.   The (amplitude) excess loss terms ($\ell_1$, $\ell_2$ and $\ell_3$) and the propagation loss $\alpha$, associated with the triplet transits are defined as (path lengths being identical, we ignore the phase): 
\begin{subequations}
\label{eq:loss_terms}
\begin{align}
\ell_1 &= \sqrt{e^{-\alpha \cdot l_{wg}} \cdot r_2^2 \cdot \mathcal{R}} \label{eq:l1_def} \\
\ell_2 &= \sqrt{e^{-\alpha \cdot l_{wg}} \cdot r_1^2 \cdot \mathcal{R}} \label{eq:l2_def} \\
\ell_3 &= \sqrt{e^{-\alpha \cdot l_{wg}} \cdot r_1 \cdot r_2 \cdot \mathcal{R}} \label{eq:l3_def} \\
\alpha &= \frac{-2\ln(a)}{L_{rt}}, \label{eq:alpha_def}
\end{align}
\end{subequations}

where $L_{rt}$ is the roundtrip length of the ring, and we make the assumption that propagation loss extracted from the rings is identical to the straight waveguide loss.

The key with bounding the value of $\mathcal{R}$ using this procedure is to ensure one accounts for all possible paths from transmitter to receiver. For the drop port device, the excess loss terms ($\ell_1$, $\ell_2$ and $\ell_3$) listed in eqn.\ref{eq:loss_terms} above correspond to the additional losses accumulated by three different permutations of the signal path comprising a triple waveguide and a half ring transit, vs the signal path corresponding to peak $\mathit{1_r}$ (cf. Fig.\ref{Fig2_zoomin}). The paths are outlined explicitly in Fig.\ref{FigA2_paths} and involve both clockwise and counterclockwise traversal of the ring on the way from transmitter to receiver IDT. These three permutations represent distinct contribution to overall loss in Eq.\ref{eq:interfaceReflection_opt} only when $r_1 \ne r_2$. Also, when $\lim_{\mathcal{R}\to0}$, the triple transit peaks disappear.

Once the valid paths (within the time/distance constraint) are identified, the cumulative effect of each path on the output signal is calculated. For each path, the cumulative phase shift is determined by summing the individual contributions from each segment (edge) in the path. The total phase shift for a given path is computed as:

\begin{equation}
\Phi = \sum_{i} \phi_i + \sum_{j} \frac{\pi}{2}\raisebox{0.35ex}{,}
\end{equation} where \( \phi_i \) is the phase shift for each propagation segment (\(\phi_i = (2\pi/\lambda)\cdot l_{prop}\)), and \( j \) is the number of couplings in the path (each coupling from waveguide to ring or ring to waveguide contributing a phase of \( \pi/2 \)). We would like to note that the $\pi/2$ phase shift is strictly true for a lossless beam splitter, but works well in practice for modelling our devices.

The total transmission power for each path is given by:

\begin{equation}
A = \mathcal{R}^{i-2} (1-\mathcal{R})^2 \prod_{j} A_j,
\end{equation} where \( A_j \) is the transmission power for \(j_{th}\) propagation segment, and \(i\) denotes the total number of port interactions in the path. We include the multiplicative factor \(\mathcal{R}^{-2}(1-\mathcal{R})^2\) to correct for the signal's injection at the L-port and extraction at the T-port, which correspond to signal transmission, rather than reflection. This compensates for the recursive pathfinding algorithm's implicit assumption that all port interactions are reflective, ensuring that the computed transmission power accurately models physical launch and detection conditions.

Each path, therefore, has an associated path length from which one can estimate the signal arrival time and the accumulated phase and propagation loss. If multiple paths converge at time \( T \), within the system timing uncertainty ($\pm$ \qty{5}{\nano\second}), their phasors are summed to account for multi-path interference:

\begin{equation}
A(T) = {\biggl|{\sum_{m \in P(T)} \sqrt{P_m} e^{j \Phi_m}}\biggl|}^2 \raisebox{-0.5ex}{,}
\end{equation} where \( P(T) \) denotes all paths whose arrival time is \( T \), \( P_m \) is the signal power, and \( \Phi_m \) is the net phase for path \( m \).

Finally, the contributions from all paths are summed to obtain the total time-domain response:

\begin{equation}
S_{21}(t) = \sum_{k} A_k \delta (t - t_k),
\label{eq:impulseResp}
\end{equation} where \( t_k \) is the time delay associated with path \( k \).

To validate the model against measured time-domain transmission data, we time-gate the signal from the launch port to the drop port to generate the template for the temporal pulse which accounts for the spectral shaping induced by the transmit and receive IDTs and the complex phase shifts at the waveguide-taper interface. The convolution of the impulse response from Eq.\ref{eq:impulseResp} with the isolated first arriving pulse (the direct transit, $1_d$) provides a model of the full temporal transmission characteristics, as noted in the main text. The comparison of the measured and simulated temporal responses demonstrates the utility of the model in capturing the complicated multipath propagation effects observed in phononic ring resonators, namely the temporal distribution of arriving acoustic pulses and the identification of interference patterns resulting from multiple propagation paths.

\begin{table}[h]
    \centering
    \renewcommand{\arraystretch}{1.2}
    \setlength{\tabcolsep}{6pt}
    \sisetup{table-number-alignment = center, table-text-alignment = center}
    \begin{tabular}{c c c} 
     \hline
     \textbf{Fit Parameter} & \textbf{Pulley-coupled Ring} & \textbf{Point-coupled Ring} \\
     \hline
     $a$ & 0.679178 & 0.818398\\
     $r_1$ & 0.683769 & 0.950522 \\
     $r_2$ & - & 0.884144 \\
     $l_{eff}$ & \qty{324.1}{\micro\meter} & \qty{203.6}{\micro\meter}\\
     $l_{idt}$ & \qty{10}{\micro\meter} & \qty{100}{\micro\meter} \\
     $\mathcal{R}$ & 0.21 & 0.08$^*$ \\ 
     \hline
     \textbf{Device Parameter} & \textbf{} & \textbf{} \\
     \hline 
     Roundtrip Length & \qty{1136}{\micro\meter} & \qty{741.77}{\micro\meter} \\
     $l_{wg}$ & \qty{269}{\micro\meter} & \qty{176.6}{\micro\meter}\\
     $v_g$ & \qty{3370}{\meter\per\second} & \qty{3535}{\meter\per\second}\\ 
     \hline
     \multicolumn{3}{l}{\footnotesize $^*$For D-Port measurement. For T-Port $\mathcal{R} \approx 0.37$} \\
     \end{tabular}
     \\[2mm]
    \caption{Model parameters for pulley-coupled rings (Fig.\ref{Fig5_graph}(b) and point-coupled rings in Fig.\ref{Fig6_graph}} 
    \label{table:A1}
    \end{table}

Table \ref{table:A1} lists the extracted model parameters for the representative point coupled and pulley coupled devices discussed in the main text. We can clearly see that the model shows that the rings approach critical coupling ($r{\approx}a$) in the pulley coupled rings from the undercoupled regime ($r>a$) in the point-coupled devices, as expected. The fitting procedure makes some simplifications, and there are two free parameters, namely $l_{idt}$ and $l_{wvg}$, that are used to align the measured and model temporal response. But once they are fixed, the rest of the model parameters can be extracted analytically and they give very good agreement with the measured data, as shown in Fig.\ref{Fig6_graph} for the drop port. As noted in the main text, we currently don't understand the reasons behind the large device-to-device variation in $\mathcal{R}$ beyond fabrication imperfections, and especially how it compares to what we can calculate using FEM simulations \cite{kcb2024piezoelectric}. 

More broadly, this computational methodology successfully captures the complex temporal dynamics of phononic ring resonators, as evidenced by Fig.\ref{Fig6_graph} fitting the drop port data. By leveraging graph-based pathfinding, phase accumulation, and attenuation modelling, this approach provides a powerful tool for analysing multi-path propagation effects in phononic circuits.

\bibliographystyle{IEEEtran}
\bibliography{IEEEabrv,References}

\end{document}